\documentclass[a4paper,11pt]{article}
\pdfoutput=1 % if your are submitting a pdflatex (i.e. if you have
\usepackage{jheppub} % for details on the use of the package, please
\usepackage[T1]{fontenc} % if needed

\usepackage{epsfig}
\usepackage{braket}
\usepackage{ulem}

\newcommand{\be}{\begin{equation}}
\newcommand{\ee}{\end{equation}}
\newcommand{\nn}{\nonumber}

\title{\boldmath Shocks and Information Exchange in de Sitter Space }

\author[a]{L. Aalsma,}
\author[b,c]{A. Cole,}
\author[b,c]{E. Morvan,}
\author[b,c]{J.P. van der Schaar,}
\author[a]{G. Shiu}

\affiliation[a]{Department of Physics, University of Wisconsin-Madison, 1150 University Ave, Madison, WI 53706, U.S.A.}
\affiliation[b]{Institute of Physics, University of Amsterdam, Science Park 904, PO Box 94485, 1090 GL Amsterdam, the Netherlands}
\affiliation[c]{Delta Institute for Theoretical Physics, Science Park 904, PO Box 94485, 1090 GL Amsterdam, the Netherlands}

\emailAdd{laalsma@wisc.edu}
\emailAdd{a.e.cole@uva.nl}
\emailAdd{e.k.morvanbenhaim@uva.nl}
\emailAdd{j.p.vanderschaar@uva.nl}
\emailAdd{shiu@physics.wisc.edu}

\abstract{We discuss some implications of recent progress in understanding the black hole information paradox for complementarity in de Sitter space. Extending recent work by two of the authors, we describe a bulk procedure that allows information expelled through the cosmological horizon to be received by an antipodal observer. Generically, this information transfer takes a scrambling time $t = H^{-1}\log(S_{\rm dS})$. We emphasize that this procedure relies crucially on selection of the Bunch-Davies vacuum state, interpreted as the thermofield double state that maximally entangles two antipodal static patches. The procedure also requires the presence of an (entangled) energy reservoir, created by the collection of Hawking modes from the cosmological horizon. We show how this procedure avoids a cloning paradox and comment on its implications.}

\begin{document} 
\maketitle
\flushbottom

\section{Introduction}
\label{sec:intro}

The main goal of this work is to study which lessons regarding complementarity in the Anti-de Sitter (AdS) eternal black hole can be transferred to de Sitter space. Since a holographic description of de Sitter space is still lacking,\footnote{Although it should be mentioned that various proposals exist, see for example \cite{Strominger:2001pn,Parikh:2002py,Alishahiha:2004md,Parikh:2004wh,Anninos:2011af,Anninos:2017eib,Anninos:2017hhn,Anninos:2018svg}.} we will rely on a pure bulk perspective involving shockwave perturbations. One important ingredient for the AdS eternal black hole is the thermofield double state, which entangles the two asymptotic regions of the maximally-extended black hole geometry. Interpreted in terms of the ER=EPR conjecture, the smoothness of the bulk geometry is a consequence of maximal entanglement between the two asymptotic (decoupled) regions \cite{Maldacena:2013xja}. In asymptotically AdS space, a convenient description is given by two entangled copies of a dual conformal field theory, in terms of which one may manipulate the dynamics and flow of information in the bulk \cite{Maldacena:2001kr}. For example, introducing a particular coupling between the two asymptotically AdS regions, or equivalently the two conformal field theories, opens up a wormhole that connects the two bulk regions and allows the exchange of information between them \cite{Gao:2016bin}. From a bulk perspective this can be understood as the introduction of negative energy shockwaves, whose backreaction connects the previously decoupled asymptotic regions \cite{Gao:2016bin}. This allows information to be transferred, and further suggests a bulk procedure involving traversable wormhole geometries for information recovery after a scrambling time avoiding the cloning paradox.  

Here we investigate to what extent these ideas also apply to de Sitter space in three or higher dimensions.\footnote{Related questions in the context of lower \cite{Maldacena:2019cbz,Cotler:2019nbi,Chen:2020tes,Hartman:2020khs,Balasubramanian:2020xqf,Sybesma:2020fxg,Aalsma:2021bit} and higher-dimensional \cite{Geng:2021wcq} models of gravity in de Sitter space have been studied.} Although the causal structure of de Sitter space is similar to that of the AdS eternal black hole, there are important differences. In particular there is no (timelike) boundary where a holographic conformal field theory perspective can provide a consistent and complete description of bulk phenomena. Even so, the ingredients introduced for information exchange in the AdS eternal black hole appear to have a valid semi-classical bulk interpretation. In particular, the entanglement described by the thermofield double state, shockwaves and (traversable) wormhole geometries seem sufficient to address some of the questions related to the information paradox. This suggests that (some of) these lessons might apply to de Sitter space as well.

Following up on recent work by two of us \cite{Aalsma:2020aib}, we 
propose a bulk procedure that allows the exchange of information between antipodal de Sitter observers after a scrambling time $t= H^{-1} \log(S_{\rm dS})$. This is consistent with de Sitter complementarity and the results of Hayden and Preskill \cite{Hayden:2007cs}. As in the AdS eternal black hole, this procedure relies crucially on selection of the Bunch-Davies state, understood as the thermofield double state, which maximally entangles the static patches of two antipodal observers in de Sitter space \cite{Goheer:2002vf}. In addition, and in contrast to the AdS eternal black hole case, we emphasize the important role of an energy reservoir, created by collecting Hawking radiation from the cosmological horizon. Manipulating this energy reservoir allows an observer in de Sitter space to interact with the cosmological horizon. In four or more dimensions the energy reservoir can be collapsed into a Schwarzschild-de Sitter black hole, opening an additional Einstein-Rosen bridge to the antipodal observer. Although we primarily present results for isotropic shockwaves in three dimensions, we expect their generalization to four or more dimensions to be straightforward. The nature of information transfer in our procedure also immediately neutralizes the threat of a potential cloning paradox. Let us emphasize that the bulk de Sitter protocol we construct applies to information exchange between anti-podal observers in de Sitter, although we will end with some remarks on de Sitter information recovery in this context. Even though most of these ingredients are not new, we believe their combination and application to de Sitter space, and the limits of such a procedure, might present a valuable first step towards understanding information recovery in cosmological spacetimes. 

We begin with a short reminder of the thermofield double state and how isotropic shockwaves in de Sitter space can in principle allow the exchange of information between two antipodal observers. After that we will review the Hayden-Preskill protocol and introduce a de Sitter bulk procedure that can describe the exchange of information between antipodal observers in a scrambling time, and how a potential cloning paradox is avoided. We end with some conclusions and remarks. 

\section{Thermofield double states and wormholes in de Sitter space}

\begin{figure}[ht]
    \centering
    \includegraphics[scale=1]{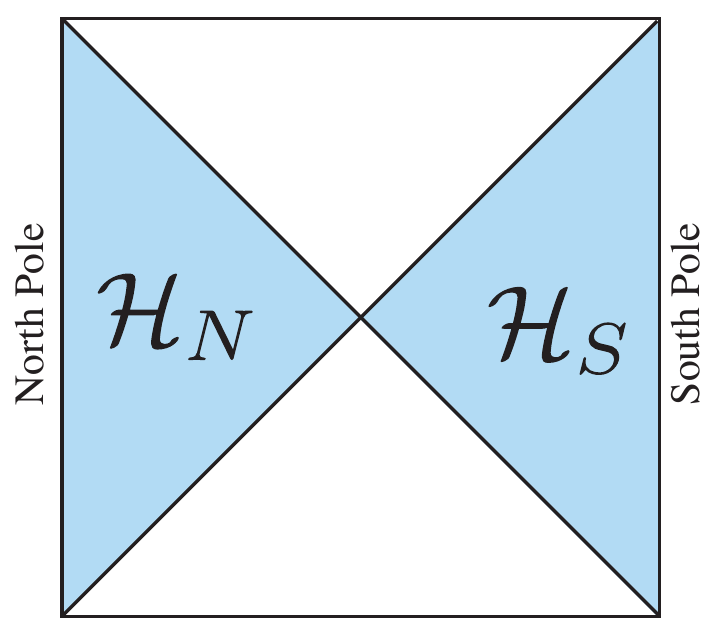}
    \caption{Two antipodal static patch regions (shaded blue). The thermofield double state with Hilbert space ${\cal H}_{\rm TFD} = {\cal H}_N\otimes{\cal H}_S$ is constructed by maximally entangling the two static patches.}
    \label{fig:dS-TFD}
\end{figure}

To begin let us remind the reader that the AdS black hole thermofield double state has a clear analogue in (semi-classical) de Sitter space \cite{Goheer:2002vf}. Consider a single static patch region of the de Sitter space, which comes equipped with a timelike Killing vector. The natural static patch Hamiltonian associated to the timelike Killing vector defines a Hilbert space, including perturbative gravitational states, with a lowest energy vacuum state that is in fact singular at the past and future horizon. To construct a regular de Sitter vacuum state one first introduces a second decoupled (antipodal) static patch region and maximally entangles it with the original static patch, setting up a thermofield double state. This of course produces the standard, regular, thermal state in each of the static patches, equivalent to the global, purified, Bunch-Davies or Euclidean vacuum. According to the ER=EPR conjecture \cite{Maldacena:2013xja} the past and future horizons connecting the interior and exterior geometry are regular and smooth as a direct consequence of the entanglement between the two static de Sitter patches, see Figure \ref{fig:dS-TFD}.

Let us remind the reader of a few important details of the thermofield double state in de Sitter space, as discussed in \cite{Maldacena:2001kr, Goheer:2002vf}, as well as the inclusion of shockwave perturbations that preserve boost invariance \cite{Giddings:2007nu}. A single static patch region, for concreteness the one anchored at the South Pole, in the standard Bunch-Davies vacuum $\ket{\rm BD}$ is described by a (mixed) thermal density matrix 

\begin{equation}
    \rho_{S}=\frac{1}{Z} \exp{(-\beta H_{S})} \, ,
    \label{thermalrho}
\end{equation} 
where $H_S$ is the Hamiltonian that generates time evolution with respect to the local timelike Killing vector.

Now consider the Hilbert space ${\cal H}_S$ of low-energy, perturbative, degrees of freedom in a single static patch at the South Pole and introduce a second identical copy ${\cal H}_N$. In the thermofield double procedure the tensor product Hilbert space ${\cal H}_{\rm TFD}={\cal H}_S \otimes {\cal H}_N$ is equipped with a Hamiltonian $H_{\rm TFD} \equiv H_S - H_N$. Geometrically the copy will be associated to the North Pole of (global) de Sitter space, and the minus sign is attributed to the opposite orientation of the timelike Killing vector at the North Pole.

The thermofield double state is given by 
\begin{equation}
    \ket{\Psi} = \frac{1}{Z^{1/2}} \sum_i \exp{\left(-\frac{1}{2}\beta E_i \right)} \, \ket{E_i}_S \, \ket{E_i}^*_N \, ,
\end{equation}
where the $^*$ refers to a CPT transformation, needed to map the states in the antipodal patch to the same orientation. With $\beta=2\pi\ell$ the inverse of the de Sitter temperature and $\ket{E_i}$ energy eigenstates, we see that the  reduced density matrix, obtained by tracing over the other pole region, indeed corresponds to the de Sitter thermal density matrix (\ref{thermalrho}). As a consequence the thermofield double purification of the static patch thermal state is equivalent to the de Sitter-symmetric Bunch-Davies (Euclidean or Hartle-Hawking) vacuum state. The vacuum expectation value of any operator ${\cal O}$ with support in just one of the static patches can be computed using the reduced density matrix: $\braket{\Psi| {\cal O}(x_S) |\Psi} = \text{Tr}_S({\cal O}(x_S)\rho_S)$, where the South Pole reduced density matrix $\rho_S$ is obtained by tracing the thermofield double state over the North Pole region $\rho_S \equiv \text{Tr}_N \ket{\Psi} \bra{\Psi}$. This also means the expectation value remains invariant under the full set of de Sitter isometries. We emphasize this point because there has been some confusion in the literature with regards to the vacuum expectation value of the stress tensor after one of the poles is traced out \cite{Markkanen:2017abw,Blumenhagen:2020doa}. We explicitly establish de Sitter invariance of the stress tensor expectation value in Appendix \ref{app:reduceddensitymatrix}. The full de Sitter invariance is simply inherited from the perturbative thermofield double construction (we do not claim this should also hold non-perturbatively). 

Importantly, the entangled thermofield double state is an eigenvector of the thermofield double Hamiltonian with vanishing eigenvalue, i.e. $H_{\rm TFD}\ket{\Psi}=0$. Correspondingly, evolution generated by this thermofield double Hamiltonian is trivial and the operator is more appropriately interpreted as providing a constraint. Indeed, in de Sitter space static time translations are related to boosts of the higher-dimensional flat embedding space; the vanishing of the thermofield double Hamiltonian simply reflects the de Sitter invariance of the state. Furthermore, to satisfy this Hamiltonian constraint, finite energy excitations (as measured by the sum of the two static patch Hamiltonians) must be introduced symmetrically, in the sense that any perturbation in the South Pole is accompanied by the same North Pole perturbation. Once the thermofield double state is prepared this way (for instance at the global timeslice $t=0$), non-trivial time evolution can be generated by the sum (instead of the difference) of the static patch Hamiltonians, introducing additional time-dependent phases in the Schr\"odinger state. 

We will use this entangled static patch set-up to argue, as for the AdS eternal black hole, that any message sent through the de Sitter horizon can be received by the antipodal observer, if a shockwave is released after at least a scrambling time. In this sense, information can be exchanged between two antipodal observers. As for the eternal AdS black holes, we will provide evidence in support of this scenario by constructing isotropic shockwave perturbations that turn the Einstein-Rosen bridge between the two static patches into a traversable wormhole geometry, connecting the two regions and allowing the information to travel smoothly to the antipodal observer. In the thermofield double state the exterior partner of an interior Hawking mode is maximally entangled by construction, and a potential de Sitter firewall paradox is avoided in exactly the same way as for the eternal black hole in AdS \cite{Maldacena:2013xja}: the original entanglement present in the thermofield double state ensures that the entanglement in any future Hawking pair is not independent, but in fact is implied by the construction. 

Clearly we will make significant use of analogies with the AdS eternal black hole scenario, but we will also emphasize some crucial differences. Primary among those is the fact that there exists a maximum energy state in de Sitter space and correspondingly a highest (isotropic and positive) energy shockwave. Having outlined the general idea let us now introduce a few more details, starting with some results of \cite{Aalsma:2020aib}. 

\subsection{Positive energy shockwaves and traversable wormholes}

\begin{figure}[ht]
    \centering
    \includegraphics[scale=1]{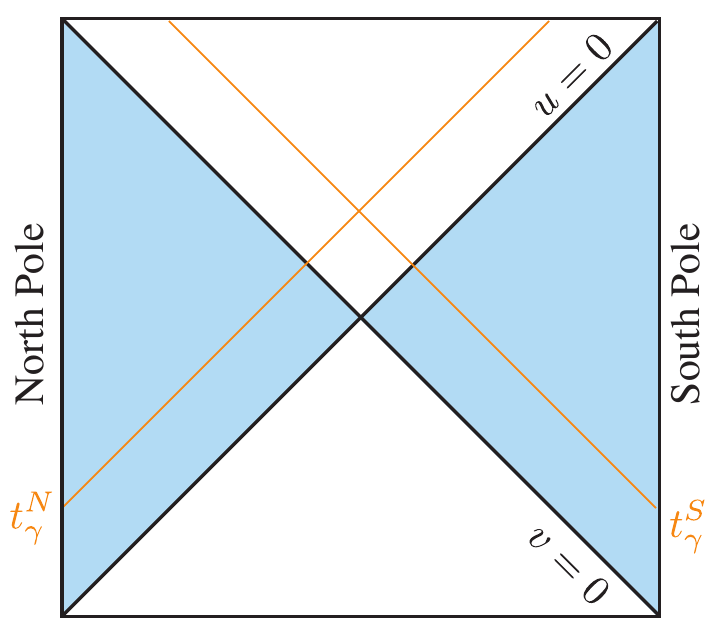}
    \caption{De Sitter Penrose diagram and global lightcone coordinates. The two static patches are shaded blue.}
    \label{fig:uv-dS}
\end{figure}

In this section we will review some facts about shockwaves and the associated traversable wormholes in de Sitter space. For these purposes it is convenient to introduce global (radial) lightcone coordinates covering $d$-dimensional de Sitter space
\begin{equation}
   ds^2 = \frac{4\ell^4}{(\ell^2-uv)^2}(-dudv)  + \ell^2\left(\frac{\ell^2+uv}{\ell^2-uv}\right)^2 d\Omega_{d-2}^2~\, . 
\end{equation}
In these coordinates $u=0$ and $v=0$ correspond to the future and past horizons, depending on the North or South Pole perspective. The global coordinates are related to planar patch coordinates covering the right half of the Penrose diagram, introducing the comoving radius $\rho$ and conformal time $\eta$, as follows
\begin{equation}
  u=\rho + \eta \quad , \quad v=\frac{\ell^2}{\rho-\eta} \, . 
\end{equation}
Planar time $\tau$ is related to conformal time as $\eta = -\ell e^{-\tau/\ell}$ such that $\eta \in (-\infty,0)$ and $\rho\geq0$. Left half planar patch coordinates can be obtained by sending $\eta\to-\eta$. In terms of static South Pole coordinates $r$ and $t$ they are expressed as
\begin{equation}
  u = - \ell e^{-t/\ell}\sqrt{\frac{\ell-r}{\ell+r}} ~, \quad v = \ell e^{t/\ell}\sqrt{\frac{\ell-r}{\ell+r}} \, .
\end{equation}
North Pole static coordinates can be obtained by sending $t\to t - i\pi \ell$, reversing the signs in the relation above. We will associate the center of the static patch to one of the poles, identifying the Bunch-Davies vacuum as the state that maximally entangles the North and South Pole static patch regions. Null rays can now be labelled by the static time when they cross the South Pole, i.e. $u(r=0)=-\ell e^{-t/\ell}$ and similarly for $v$. Note that when evolving with the sum of the static patch Hamiltonians the roles (future or past horizon) of $u$ and $v$ are interchanged when moving from the North to the South Pole static patch, see Figure \ref{fig:uv-dS}. Also observe that static time translations correspond to a boost of the global lightcone coordinates (acting as rescalings). 

Shockwave geometries can be derived by boosting Schwarzschild-de Sitter solutions \cite{Hotta:1992qy,Hotta:1992wb,Sfetsos:1994xa}. Here we will restrict ourselves to the three-dimensional case, where the relevant geometry corresponds to a particle with mass $m$, bounded by $m<\frac{1}{8G_N}$, at the center of the static patch. The final result for a shockwave in global lightcone coordinates travelling along the $v=0$ horizon is \cite{Aalsma:2020aib}
\begin{equation} \label{eq:horizonshockwave}
    ds^2 = \frac{4\ell^4}{(\ell^2-uv)^2}(-dudv) - 4\alpha \delta(v) dv^2 + \ell^2\left(\frac{\ell^2+uv}{\ell^2-uv}\right)^2 d\phi^2 \, . 
\end{equation}
Starting from the particle geometry the parameter $\alpha=2\pi G_N \ell p$ is related to the energy $p$ obtained in the infinite boost limit $\beta \rightarrow 1$ of the particle in de Sitter solution while at the same time taking $m\to 0$
\begin{equation}
    p = \frac{m}{\sqrt{1-\beta^2}}  \, ,
\end{equation}
resulting in a shockwave travelling along the $v=0$ horizon with finite energy $p$. The stress tensor of this shockwave is given by
\be \label{eq:shock}
T_{vv} = \frac{\alpha}{4\pi G_N\ell^2}\delta(v) \, .
\ee
The null energy condition imposes $\alpha\geq0$. Geodesics crossing the shockwave at $v=0$ orthogonally experience a negative shift in the $u$ direction. Boosts rescale the null coordinate $v$, so the generalization to isotropic shockwaves not traveling along the past horizon can be made by identifying the energy $p$ as the energy of an isotropic shockwave as it crosses the origin. Explicitly, we will construct a Vaidya geometry describing a spherical matter shell expanding in the static patch. This leads to a geometry in which geodesics crossing the shell close to the horizon experience the same shift as for the infinite boost geometry.\footnote{Analogous off-horizon isotropic shockwave geometries have been constructed in \cite{Verheijden:2018}.}

We will focus on a shell that is emitted from the South Pole and define two sets of null coordinates $(\tilde U,\tilde V)$ and $(U,V)$ that respectively describe the geometry to the left (exterior) and right (interior) of the shell. The interior geometry is empty de Sitter space and the exterior geometry Schwarzschild-de Sitter with energy parameter $\mathcal{E}$. We can fix the relation between these coordinates by imposing that the radius of the $S^1$ is continuous across the shell and requiring that $t$ has no discontinuity at the pole. Focusing on three dimensions, we then have
\begin{align}
U &= t+r_* ~, \quad V = t-r_*~, \\
\tilde U &= t+\tilde r_* ~, \quad \tilde V = t-\tilde r_* \nn ~.
\end{align}
with the tortoise coordinates $r_*,\tilde r_*$ defined as
\begin{align}
r_* &= \int_0^r \mathrm{d}r'\left(1-r^2/\ell^2\right)^{-1} = \ell\, \text{arctanh}(r/\ell) ~, \\
\tilde r_* &= \int_0^r \mathrm{d}r'\left(1-8G_N \mathcal{E} -r^2/\ell^2\right)^{-1} = \frac{\ell}{1-8G_N \mathcal{E}} \text{arctanh}\left(\frac{r}{\ell\sqrt{1-8G_N \mathcal{E}}}\right) \nn ~.
\end{align}
We will be interested in the behaviour of null geodesics in this geometry. A geodesic travelling orthogonal to the shell (at constant $\tilde U$) obeys $\mathrm{d}\tilde V +2\mathrm{d}\tilde r_* = 0$. This expression is valid for a geodesic to the left of the shell and to obtain the expression to the right of the shell we remove tildes from the coordinates. Expressed in terms of the $u=-\ell e^{-U/\ell}$ coordinate, the discontinuity in the $u$ coordinate when crossing the shell is
\be
\tilde u - u= - \ell e^{-2\tilde r_*/\ell} + \ell e^{-2r_*/\ell} ~.
\ee
Evaluated at the horizon and for small $G_N \mathcal{E}$ this reduces to a constant shift $\alpha = \tilde u - u \approx 2G_N \mathcal{E} \ell$. The shift in the $u$ coordinate in this near-horizon, small $G_N\mathcal{E}$, limit of an isotropic shockwave is equal to the shift $\alpha$ in the infinite boost limit \eqref{eq:horizonshockwave} which relates $\pi p= {\cal E}$. The metric of the shell is described by the following Vaidya geometry
\be
\mathrm{d}s^2 = -\left(1-8G_N \mathcal{E} \theta(V_0-V) - r^2/\ell^2\right)\mathrm{d}V^2 -2\mathrm{d}V\mathrm{d}r + r^2\mathrm{d}\phi^2 ~,
\ee
which solves Einstein's equations with a stress tensor
\be
T_{VV} = \frac{\mathcal{E}}{2\pi r}\delta(V_0-V) ~.
\ee

Besides the fact that these are positive energy shockwaves, as opposed to the negative energy shockwaves introduced in AdS, another important difference with respect to AdS is that we will interpret the presence of positive energy as a result of collecting (s-wave) Hawking modes from the cosmological horizon, maintaining the thermofield double constraint. This produces a spherically symmetric mass configuration of the same mass at the two de Sitter poles, which we assume can be appropriately stored and consequently manipulated into outgoing positive energy shockwaves. This forces us to introduce a cut-off on the energy parameter $\mathcal{E}$ in three dimensions, as a direct consequence of the bound on the mass of a spherically symmetric configuration in de Sitter space, i.e. $\mathcal{E} \leq 1/8 G_N \propto M_p$. This bound can in fact be generalized to higher dimensions in terms of the de Sitter entropy and horizon length scale $\ell$, up to some dimension-dependent order one factors that we will collect in a coefficient $c_d$, 
\begin{equation}
    \mathcal{E} \leq c_d \, \frac{S_{\rm dS}}{4\pi \ell} \, .
    \label{ESbound}
\end{equation}
In three dimensions $c_3=1$, whereas in four dimensions it equals $c_4=4/(3\sqrt{3})$. In three dimensions this equals the Planck mass, but in four dimensions it is related to the largest (Nariai) black hole that fits in de Sitter space $M=\ell/(3\sqrt{3}G_N)$, and therefore bigger than $M_p$ by a factor of order $\ell M_p$.\footnote{Classical de Sitter black holes obviously require the mass to be bigger than the Planck mass in order for the Compton wavelength to fit inside the Schwarzschild radius. As a consequence, black hole solutions do not exist in three dimensions.} In addition, due to the finite volume of the static patch (or the de Sitter temperature), there also exists a lower bound on the energy of an excitation in de Sitter space, namely $\mathcal{E} \geq 1/ (2\pi \ell)$. Note that the upper bound on the energy (\ref{ESbound}) implies that the lower bound can be associated to an order one unit of entropy in de Sitter space. Defining $N_s$ as the energy of the shockwave in units of the de Sitter temperature $N_s \equiv \mathcal{E}(2\pi\ell)$, this allows one to write $2 \leq 2N_s \leq c_d S_{\rm dS}$, which will turn out to be a rather useful parameterization.

For our purposes we will prepare an initial, boost symmetric, de Sitter thermofield double state, including an energy reservoir at both poles at $t=0$. Both energy reservoirs are then turned into a pair of shockwaves travelling along $u=\ell$ and/or $v=\ell$. As we saw, in three dimensions null geodesics crossing one of these shockwaves will then be translated along the (orthogonal) lightcone coordinate ($\Delta v$ for a $u=\ell$ shockwave) by an amount $\alpha = 2 G_N \mathcal{E} \, \ell$, as long as the null geodesics are (relatively) close to the future horizon. This expression for the shift can be rewritten using the three-dimensional de Sitter entropy $S_{\rm dS}=(2\pi \ell)/4G_N$ and $N_s$ as
\begin{equation}
    \alpha = \left( \frac{N_s}{2 S_{\rm dS}} \right) \, \ell  ~,
\label{shift0}
\end{equation}
which we again expect to generalize directly to four or higher dimensions, possibly introducing another order one factor. In three dimensions, using that $2N_s \leq S_{\rm dS}$, the maximal shift is  of order the de Sitter length, $\frac{1}{4} \ell$ to be precise. For the minimal energy $N_s = 1$, the shift is instead of the order of a Planck length. Note that this range is obviously related to the change in the de Sitter horizon due to collecting or expelling isotropic energy from the static patch, which is a Planck length for energy changes of order $1/(2 \pi \ell)$, and a de Sitter length for energy changes of order $S_{\rm dS}$. This shift of the de Sitter horizon can also be associated to the size of a traversable wormhole, opening up an information channel between the two antipodal observers.\footnote{Deviations from a constant shift are expected to set in at second order in the lightcone coordinate, which we will ignore throughout this paper.}

Although we will specifically be considering three dimensions, as we tried to emphasize by writing the energy bound and geodesic shift in terms of the de Sitter entropy and the length scale $\ell$, our results should readily generalize to four (and higher) dimensions.  

\subsection{De Sitter wormholes and a bound on the flow of information}

The shift of de Sitter null geodesics as they cross an isotropic shockwave shell effectively extends the causal conformal diagram \cite{Gao:2000ga}, implying that the North and South Pole static patches can now exchange information. As was elegantly explained in the context of the eternal black hole \cite{Gao:2016bin,Maldacena:2017axo}, the maximally entangled but causally disconnected asymptotic AdS regions can be brought into causal contact by introducing a symmetric coupling in the holographically dual conformal field theory. This perturbs the thermofield double state and relates the gravitational exchange of (quantum) information to teleportation in the holographic conformal field theory. From the AdS bulk perspective this corresponds to the introduction of a symmetric pair of negative energy shockwaves entering the bulk, effectively opening up a central wormhole region, making the Einstein-Rosen bridge traversable for some period. Sending negative energy into the black hole exposes (part of) the interior geometry and allows information to leak out. 

In a series of thought experiments \cite{Maldacena:2017axo} the combination of maximal entanglement with the introduction of negative energy shockwave perturbations was enough to explain how a firewall and the cloning of information is avoided, in support of black hole complementarity. Importantly, for a one-sided black hole in AdS space it was proposed that the thermofield double state can be well approximated by collapsing a large number of Hawking modes, of the order of the black hole entropy, into a black hole, leading to two black holes that are maximally entangled and that can effectively be understood as the two (entangled) sides of the eternal AdS black hole \cite{Maldacena:2013xja}. This allows for an interpretation of information recovery in terms of a bulk gravitational description of the Hayden-Preskill teleportation protocol, allowing information thrown into the original AdS black hole to be recovered after a scrambling time from the collapsed Hawking radiation. In the bulk gravitational description this is elegantly understood as the information smoothly traveling through a wormhole connecting the two black holes.  

\begin{figure}[ht]
    \centering
    \includegraphics[scale=1]{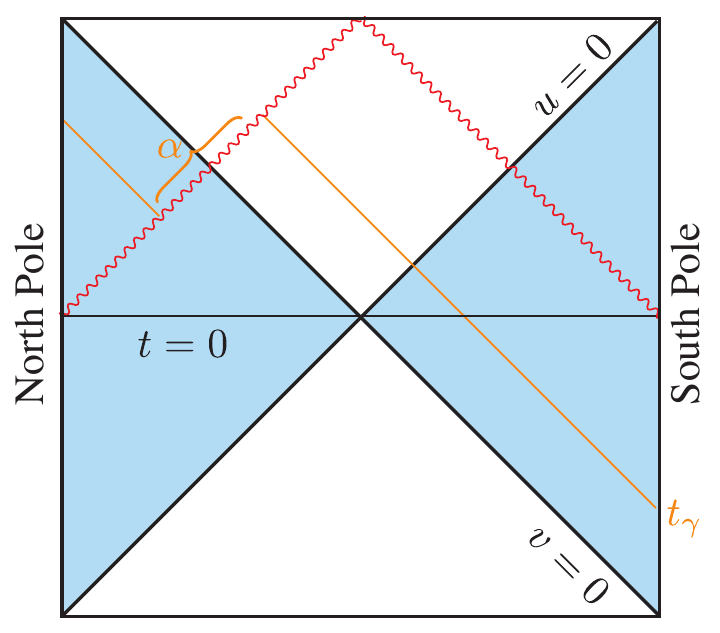}
    \caption{A pair of shockwaves in de Sitter space and a shifted lightray}
    \label{dS-shockwave}
\end{figure}

We propose that the situation for two antipodal observers in de Sitter space should be rather similar, from a pure bulk perspective. To start, the de Sitter symmetric Bunch-Davies vacuum is equivalent to the thermofield double state that maximally entangles all pairs of antipodal static patch regions, connected by an Einstein-Rosen bridge, producing a smooth global de Sitter geometry. This de Sitter Einstein-Rosen bridge can be turned into a (boost symmetric) traversable wormhole geometry by introducing a symmetric pair of {\it positive} energy isotropic null shells injected at $t=0$ at the centers of two antipodal static patches (the North and South Pole respectively). This effectively moves the location of the cosmological de Sitter horizon, extending the causal structure, allowing the exchange of information between two antipodal observers.\footnote{This idea has also been explored by E. Verheijden in her thesis \cite{Verheijden:2018} and a set of unpublished notes.} See Figure \ref{dS-shockwave}. However, in contrast to the AdS eternal black hole, we will have little to say about how to promote this exchange protocol to an information recovery procedure for a single static observer.  

As for the eternal black hole in AdS we can derive a simple bound on the information that can be sent through the wormhole, generalizing the result in \cite{Aalsma:2020aib}, employing the bounds on the energy of de Sitter shockwaves that we introduced. First we remind the reader that in three dimensions the maximal shift $\alpha$ on a (null) geodesic crossing the shockwave is of the order of the Hubble length, when $2N_s = S_{\rm dS}$, where $N_s$ corresponds to the energy in units of the de Sitter temperature. This assumes the information propagating along the geodesic is not itself backreacting significantly on the geometry, as opposed to the energy in the isotropic shockwave shell. Self-consistency therefore suggests that the total energy in the probe trajectory $N_{\rm bit} \epsilon_{\rm bit}$, where $\epsilon_{\rm bit}$ corresponds to the energy of a single bit, is small in comparison to $\mathcal{E}$, i.e. $N_{\rm bit} \epsilon_{\rm bit} < \mathcal{E} \leq S_{\rm dS}/(4 \pi \ell)$. Because the smallest energy of a single bit is $1/(2 \pi \ell)$, this immediately tells us that the total number of bits $N_{\rm bit}$ is smaller than $N_s$, which is bounded from above by the de Sitter entropy 
\begin{equation}
    N_{\rm bit} < N_s \leq \frac{1}{2} S_{\rm dS} \, . 
\end{equation}    
This is an immediate consequence of the probe approximation in combination with a highest energy state in de Sitter. In addition, the uncertainty principle tells us that the information in the probe trajectory has a spread equal to $1/\epsilon_{\rm bit}$. For the information to go through the wormhole this spread should be contained within the size of the wormhole, which is measured by the shift $\alpha$ (\ref{shift0})
\begin{equation}
   \epsilon_{\rm bit} \, \ell > \frac{\ell}{\alpha} = \frac{2 S_{\rm dS}}{N_s} \, . 
\end{equation}
Using the threshold value for $\epsilon_{\rm bit}$ to fit in the wormhole ($\epsilon_{\rm bit} = 2S_{\rm dS}/(\ell N_s) $) one then ends up with the following constraint on the number of bits $N_{\rm bit}$ that are consistent with the probe approximation $N_{\rm bit} \epsilon_{\rm bit} < \mathcal{E}$
\begin{equation} \label{eq:probeconstraint}
    N_{\rm bit} < \frac{N_s^2}{4 \pi S_{\rm dS}} \, . 
\end{equation}
To be able to transfer a non-vanishing number of bits $N_{\rm bit}$ will require $N_s \gtrsim \sqrt{S_{\rm dS}}$. The quadratic scaling with $N_s$ is a direct consequence of the fact that the minimum energy of a single bit scales as $S_{\rm dS}/N_s$, implying that for $N_{\rm bit} \lesssim N_s$ the probe condition cannot be satisfied. If one would instead just impose the maximal de Sitter energy bound $N_{\rm bit} \epsilon_{\rm bit} < S_{\rm dS}/(4 \pi \ell) \propto M_p$, one would end up with $N_{\rm bit} \lesssim N_s$.

Independent of the particular energy constraint, for the maximal energy shockwave one reproduces $N_{\rm bit}<N_s \leq \frac{S_{\rm dS}}{2}$. The number of bits that can be sent through the wormhole is always smaller than the number of `classical bits' $N_s$ represented by the shockwave, which is always smaller than the de Sitter entropy. The derivation presented here relies on the details of the probe approximation, as well as the size of the wormhole shift parameter $\alpha$, and allows for general shockwave energy $\mathcal{E} \propto N_s$, whereas the result in \cite{Aalsma:2020aib} assumed $N_s=1$, i.e. the minimal energy shockwave. In the latter case, one is forced to introduce $K$ species to relax the bound to be of the order of the de Sitter entropy. 

We can summarize this result as follows. The emission of an isotropic shockwave shell expels energy from the static patch region, increasing the cosmological horizon and gravitational entropy accordingly. The shockwave can maximally increase the entropy represented by the cosmological horizon to that of empty de Sitter space. An amount of information (on a null geodesic) crossing the shockwave and entering the static patch region will decrease the de Sitter entropy ever so slightly, due to its own backreaction, but by a lot less than the increase produced by the outgoing positive energy shockwave itself, if the probe approximation is satisfied. The amount of information that can be transferred using shockwaves is therefore always significantly smaller than the number of `classical bits' $N_s$ required to set-up the wormhole configuration, consistent with a quantum teleportation protocol, as we will review shortly. 

\section{Information transfer and complementarity in de Sitter space}

We will present a short summary of the status of complementarity in de Sitter space, briefly reminding the reader about the Page and scrambling times, as well as how the firewall and cloning paradoxes are avoided in the AdS eternal black hole context \cite{Maldacena:2017axo, Hayden:2007cs, Maldacena:2013xja, Danielsson:2002td, Parikh:2008iu}. To begin let us remind the reader that a consistent unitary quantum mechanical description of a causal region with finite de Sitter entropy, respecting the de Sitter symmetries, is in fact known to be impossible beyond the recurrence time \cite{Goheer:2002vf}. After this time, the system starts to explore extreme regions of phase space (in e-folds this is $N\equiv H\tau \propto e^{S_{\rm dS}}$). This suggests that the de Sitter symmetries are broken when gravitational interactions are taken into account (for finite $M_p$, resulting in a finite gravitational entropy). Even so, we will be working under the assumption that this should not significantly affect phenomena at much shorter time-scales, like the Page or scrambling time ($N\propto S_{\rm dS}$ or $\log{(S_{\rm dS})}$ respectively), which is what we will be interested in. On these shorter timescales we will use (approximate) de Sitter symmetry and assume that the complementary experiences of all free-falling observers can be described by the same (unitary) quantum description. This must consequently mean that the different experiences of free-falling observers of a particular (horizon-crossing) event should not be in conflict with basic tenets of quantum mechanics. Specifically, it should not present an observable violation of the no-cloning principle.

We will present evidence that information can be exchanged between antipodal observers in de Sitter space in a scrambling time. Indeed, assuming a maximally entangled state, finite de Sitter entropy, and unitary evolution, it should in principle be possible to recover information after a scrambling time, as was shown convincingly using methods from quantum information theory \cite{Hayden:2007cs}.\footnote{This assumes the observer at has full access to and can manipulate the large $\sim e^{S_{\rm dS}}$ number of states, which in practice might be difficult.} Before this basic result was discovered, in the context of de Sitter space the much longer Page time had been suggested as the time-scale before a single bit of information could be distilled \cite{Page:1993df}.\footnote{And importantly, one would have to collect all the radiated Hawking quanta for at least the Page time. Note that \cite{Aalsma:2021bit} considered a different setup in de Sitter space where one does have to wait for a Page time to recover information.} If that were the case, observer complementarity in de Sitter space is safe, either as a result of the maximum energy that can be collected in a finite size de Sitter static patch (for $d>2$) \cite{Parikh:2008iu}, or, if one exits the de Sitter phase into a flat FLRW cosmology, because the time for the information to reappear into the increasing particle horizon is longer than the de Sitter recurrence time \cite{Danielsson:2002td}.

Assuming the asymptotic future geometry is (empty) de Sitter, if one instead only needs to wait a scrambling time to collect the relevant information from just a few cosmological Hawking modes, the no-cloning principle should again be in jeopardy. If the Hayden-Preskill protocol applies in de Sitter, an observer could throw a single (secret) bit of information through the horizon, and then after waiting a scrambling time collect a few bits of information from the Hawking radiation to decode the information. The observer could then decide to jump in and potentially see the same bit twice. The absence of a spacelike singularity in de Sitter space would suggest there is no obstruction to observe the same bit twice, corresponding to a violation of the no-cloning principle, and thus (observer) complementarity. 

More recently, in the context of eternal AdS black holes, it was pointed out that one does not need to rely on obstructions to observe the cloned information \cite{Maldacena:2013xja, Maldacena:2017axo}. Instead, since interior and exterior observables are not commuting in the thermofield double state, the information is never copied in the first place. If one records the information in the exterior, the information in the interior is removed, as a consequence of the entanglement, and vice-versa. The absence of a second copy, instead of the practical impossibility to observe it, is clearly a more satisfactory scenario, which one would then expect to be operating in de Sitter space as well. Before discussing the details of the particular gravitational procedure, let us remind the reader of the Hayden-Preskill protocol in general and its application to de Sitter space in particular. 

\subsection{The Hayden-Preskill protocol in de Sitter space}

\begin{figure}[h]
\centering
\includegraphics[scale=.5]{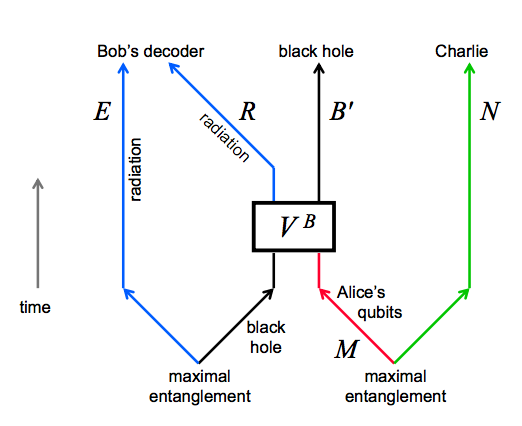}
\caption{Quantum circuit corresponding to the Hayden-Preskill protocol. Figure from \cite{Hayden:2007cs}.}
\label{fig:HP_circuit}
\end{figure}

Before we discuss information exchange in de Sitter space, let us first briefly remind the reader about some of the basics of quantum information theory needed for the Hayden-Preskill protocol. First of all, when Bob collects Hawking radiation in order to retrieve information that Alice threw across the horizon, when do we say that Bob has retrieved that information? From an information theoretic perspective, what needs to happen is that the information carried by Alice's qubits is transferred to a system that Bob has access to. For example, given a quantum state consisting of two qubits
\be
\ket{\psi}=\frac1{\sqrt{|A|}}\sum_i\ket{i}_A\ket{0}_B ~,
\ee
we can consider an operation that moves information from qubit $A$ to $B$:
\be
\ket{\psi} \to \ket{\psi'}= \frac1{\sqrt{|A|}}\sum_i\ket{0}_A\ket{i}_B ~.
\ee
How can we check that this transfer was successful? We can couple the two-qubit state to an auxiliary system $C$ that is maximally entangled with $A$.
\be
\ket{\Psi} = \frac1{\sqrt{|A|}}\sum_i\ket{i}_A\ket{0}_B\ket{i}_C ~.
\ee
Performing the same information-transferring operation now leads to
\be
\ket{\Psi} \to \ket{\Psi'}= \frac1{\sqrt{|A|}}\sum_i\ket{0}_A\ket{i}_B\ket{i}_C ~.
\ee
This means that the purification of $C$ has been moved from $A$ to $B$. We can therefore check if the information transfer was successful by comparing the reduced density matrices $\rho_{AC}$ with $\rho_A\otimes\rho_C$. Before the transfer, $\rho_{AC}$ is a maximally entangled state, but after the transfer $C$ is no longer entangled with $A$. In terms of the trace distance between $\rho_{AC}$ and $\rho_A\otimes \rho_C$, information transfer was successful when\footnote{In practice, the dynamical mechanism that transfers information, such as a black hole that emits information as Hawking radiation, might involve additional unitary operators. This means that to recover information, one first has to act with the Hermitian conjugates of the unitaries. This should be possible in principle, but might be an extremely difficult operation in practice.}
\be
||\rho_{AC}-\rho_A\otimes\rho_C|| \ll 1 ~.
\ee
Hayden and Preskill applied these ideas to a single-sided evaporating black hole \cite{Hayden:2007cs}. They considered the situation where Alice has some information in a system $M$ that is maximally entangled with Charlie's reference system $N$. She then waits for the black hole to evaporate for a Page time until a sufficiently large system $E$ exists that consists of radiation that is maximally entangled with the black hole. Alice then throws system $M$ into the black hole (which we denote by system $B$). The absorption of $M$ into $B$ is modeled by a unitary $V^B$ that acts on the combined $B+M$ system, see Figure \ref{fig:HP_circuit}.
If Bob has access to system $E$ and collects the new radiation (system $R$), when does he recover $M$? According to our previous discussion, this happens when the purification of $N$ moves from $B'$ to $R$. Thus, Bob has successfully recovered $M$ when
\be
||\rho_{B'N}-\rho_{B'}\otimes\rho_{N}|| \ll 1 ~.
\ee
The trace norm is given by (integrating uniformly over all unitaries with respect to the Haar measure) \cite{Hayden:2007cs}
\be
\int dV^B ||\rho_{B'N}-\rho_{B'}\otimes\rho_{N}|| \leq \frac{|N|^2}{|R|^2} = \frac{2^{2k}}{2^{2(k+c)}} = 2^{-2c} ~.
\ee
Here $k$ is the number of qubits in Alice's message and $c$ are the additional qubits in the radiation. Thus, the message comes out of the black hole very quickly, namely as soon as a few more qubits have been emitted by the black hole than were in the original message. 

We can perform a similar procedure in de Sitter space. However, an important difference with the single-sided evaporating black hole case is that in de Sitter space the natural (Bunch-Davies) vacuum is maximally entangled with respect to any two antipodal observers to begin with and does not need to be carefully prepared by collecting Hawking radiation for at least a Page time. The Bunch-Davies state corresponds to a stable thermal equilibrium \cite{Gibbons:1977} and the de Sitter horizon is therefore not `evaporating', as for the black holes in flat space.\footnote{This is not necessarily true if the de Sitter isometries are spontaneously broken, as has been suggested \cite{Goheer:2002vf}, but as we argued before this is expected to occur at a time scale much longer than the scrambling time.} Making use of the maximal entanglement between two antipodal observers, we will now argue that the Hayden-Preskill protocol still applies in much the same way as for black holes, where Alice and Bob are now associated to two antipodal observers. The fact that Alice and Bob need to be identified with two antipodal observers in de Sitter space is important, because it obstructs a straightforward extrapolation of the teleportation procedure to an information recovery protocol in a single static patch.  

\begin{figure}[h]
\centering
\includegraphics[scale=1]{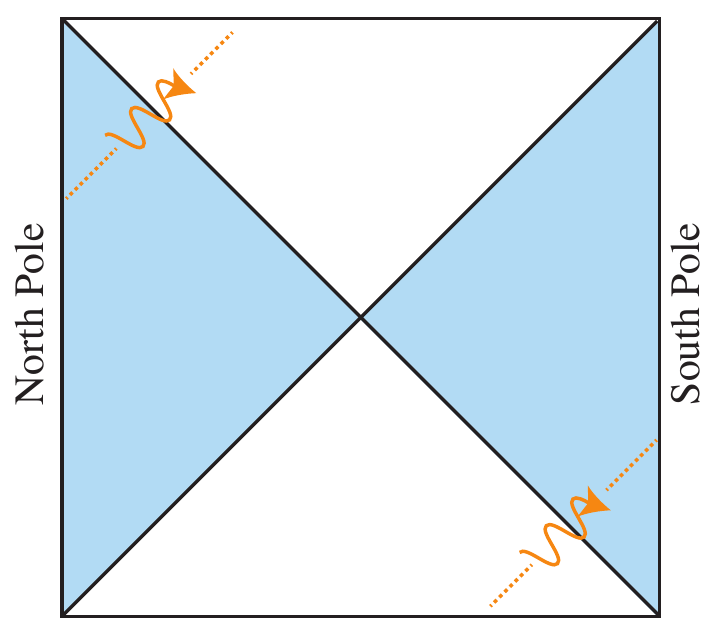}
\caption{In the thermofield double state, for each mode there is a maximally entangled mirror mode across the horizon.}
\label{fig:dS_entangled}
\end{figure}

Let us say that Bob lives at the South Pole of de Sitter space and has access to and can manipulate system $E$, consisting of all radiation in the static patch. In the thermofield double state, for every mode a maximally entangled partner exists beyond Bob's horizon, see Figure \ref{fig:dS_entangled}.
Now, Alice throws system $M$ (which is maximally entangled with reference system $N$) from the North Pole through the future de Sitter horizon. After this happens, Bob starts collecting additional Hawking radiation. From a quantum information perspective, the situation is now equivalent to the (single-sided) evaporating black hole case. If the purification of $N$ transfers to the new radiation (system $R$), Bob has successfully obtained Alice's information and Hayden and Preskill argue this happens when Bob has collected a few more qubits than were in the original message. As the time to collect (a few) Hawking modes is much shorter than the scrambling time, which measures the time for system $M$ to thermalize at Bob's horizon, the duration to transfer the information is set by the scrambling time.  

Assuming a complete, unitary, description of the de Sitter static patch exists, the above (teleportation) protocol suggests that two antipodal observers should at least be able to exchange information in roughly a scrambling time. Doing so in practice might be difficult. For example, one limiting aspect of decoding information in de Sitter space is the fact that an observer only has access to a finite amount of resources in his static patch. Despite these practical challenges, we will describe a bulk gravitational protocol using shockwaves to argue that, as in \cite{Maldacena:2017axo}, information can be exchanged successfully in a scrambling time.

\subsection{Transferring information between antipodal observers}

We will describe a set-up using isotropic shockwave shells that allows for the exchange of information between two antipodal observers in de Sitter space. This is analogous to transferring information between the two asymptotic regions of the eternal AdS black hole \cite{Maldacena:2017axo}. For a message sent from the South Pole, say from Bob, to reach an observer, say Alice, at the North Pole, the probe message has to be sent by Bob some time $t_\gamma<0$ before setting up entangled state perturbed by a shockwave perturbation. We can parameterize the null curve along which the message travels as $v_\gamma=\ell e^{t_\gamma/\ell}$ and the shockwave sent in from the South Pole at $t=0$ as $v_0=\ell$. The shockwave shell sent in from the North Pole instead travels along the constant $u_0=\ell$ null curve. As the message intersects the $u_0$ shockwave shell it will experience a time advance, or $\Delta v$ shift, measured by the parameter $\alpha$. The set-up is displayed in Figure \ref{dS-shockwave}. In three dimensions the shift can be usefully expressed as $\alpha= \frac{N_s}{2 S_{\rm dS}} \ell $ (\ref{shift0}), where we remind the reader that $N_s = \mathcal{E} (2\pi \ell)$ is the total energy of the isotropic shockwave in units of the de Sitter temperature. We would like the message to end up in the North Pole static patch, which implies that the final shifted null curve travels along a fixed negative $v'_\gamma<0$, i.e. we demand that 
\begin{equation}
    v'_\gamma = v_\gamma - \alpha = \ell \left( e^{t_\gamma/\ell} - \frac{N_s}{2 S_{\rm dS}} \right) < 0 \, . 
\label{transfercondition}
\end{equation}
Since $2N_s < S_{\rm dS}$ the number of e-folds before the shockwave is released at $t=0$ $N_\gamma \equiv -t_\gamma/\ell$, should be sufficiently large, giving 
\begin{equation}
    N_\gamma \geq  \log\left(\frac{2 S_{\rm dS}}{N_s}\right) \, .  
\end{equation}
For any fixed value of $N_s$ one observes that $N_\gamma$ scales as $\log{(S_{\rm dS})}$, which is commonly referred to as the scrambling time. Imposing the constraint \eqref{eq:probeconstraint} for a single bit we find $N_s \sim \sqrt{S_{\rm dS}}$ which will reduce this number of e-folds by half, but maintains the logarithmic scaling with the de Sitter entropy. As a curiosity, we note that in four dimensions $N \leq \frac12\log(S_{\rm dS})$ coincides with the TCC bound \cite{Bedroya:2019snp,Bedroya:2020rmd}.

By looking at Figure \ref{dS-shockwave} we can understand why this happens: if the message is not sent early enough, or equivalently not close enough to the past horizon at the South Pole, the null curve is not close enough to the future horizon of the North Pole static patch and the shift will be unable to map the message into the North Pole static patch region. We conclude that Bob at the de Sitter North Pole can actually get a hold of Alice's message by perturbing the Bunch-Davies state at $t=0$ with a shockwave with energy $\mathcal{E}=N_s/(2 \pi \ell)$. As in the AdS eternal black hole case, we will interpret this exchange of information between antipodal observers in de Sitter space as a bulk gravitational representation of the Hayden-Preskill teleportation protocol, allowing a message to be exchanged in a scrambling time. We will soon present a more precise estimate for the time it takes for the message to smoothly arrive through the wormhole at the North pole, but it is anticipated that it naturally scales as the scrambling time.

A priori, the introduction of a symmetric pair of shockwave shells seems unnecessary, since only the shockwave released at the North Pole affects the message sent from the South Pole. Because constant time slices in de Siter space are compact, a Gauss constraint in de Sitter space will ensure that the collected energy at the two antipodal regions is the same, also guaranteeing maximal (and boost symmetric) entanglement between the two poles. It does not exclude the possibility to consequently disperse this energy asymmetrically, in just one of the antipodal regions, but this will necessarily result in a final state that cannot be interpreted as (or generalized to) a smooth traversable wormhole geometry, open for a finite amount of proper time \cite{Freivogel:2019whb}, connecting two empty de Sitter static patches. So the symmetry requirement can be understood as the (future) boundary condition of interest, although it would certainly be interesting to study the details of an asymmetric isotropic shockwave setup, including its potential holographic interpretation in terms of quantum teleportation. 

\begin{figure}[ht]
\centering
\includegraphics[scale=1]{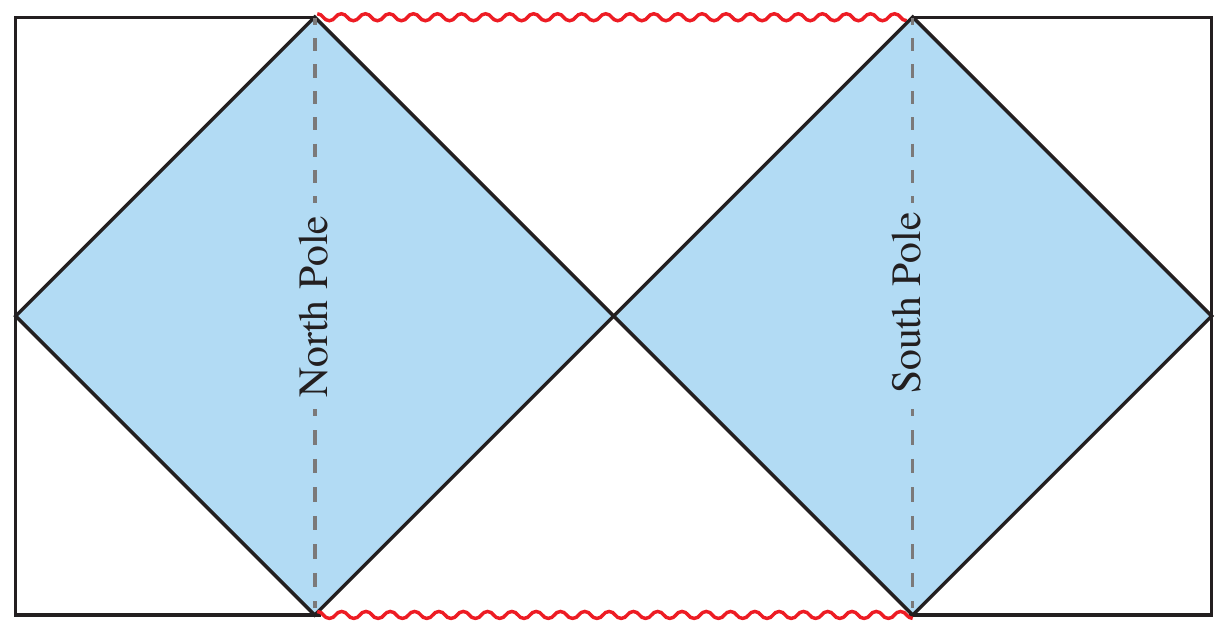}
\caption{The Schwarzschild-de Sitter causal diagram showing the shared exterior, as well as a shared black hole interior, between two (previously) antipodal observers. As the black hole is removed the interior geometry of the black hole is replaced with a de Sitter exterior geometry.}
\label{fig:dS-S-Penrose}
\end{figure}

As we stressed, due to the Gauss constraint in de Sitter, Bob and Alice should collect the same energy at the de Sitter poles. The only states that Bob and Alice have access to are the Hawking modes emitted from the de Sitter horizon, which allows them to collect energy at an average rate of one unit of $1/\ell$ per e-fold. In four dimensions or higher, after having collected a large (enough) number of Hawking modes these modes can be collapsed into a black hole. Since these cosmological Hawking modes are entangled with the antipodal region, in the spirit of the ER=EPR conjecture, we expect the corresponding black hole interior to be shared between the two antipodal static patches. In other words, the corresponding Schwarzschild-de Sitter geometry then connects the two antipodal regions via an additional Einstein-Rosen bridge, passing through the (black hole) center of the respective static patch region, see Figure \ref{fig:dS-S-Penrose}.

According to the ER=EPR conjecture the above observations can be extended beyond the semi-classical gravity regime and apply just as well to (entangled) matter configurations that have not necessarily collapsed into black holes. Collecting energy from entangled Hawking modes is in that sense the de Sitter analogue of the thermofield double construction for a physical (AdS) black hole, with the important difference that there seems to be no need to collect Hawking radiation for a Page time. Once Bob and Alice have prepared such a finite energy, maximally entangled, Schwarzschild-de Sitter state, they can now interact with the various horizons to produce a traversable wormhole geometry and exchange information at a rate set by the scrambling time. As already alluded to, we will assume a symmetric configuration, but as for the eternal AdS black hole, a bulk description in terms of shockwaves does suggest it should be possible to deform this setup to (final) states that are not symmetric and, as a consequence, not maximally entangled. 

\subsection{Preparing the state and resolving the cloning paradox}
Now that we have set up a finite energy, maximally entangled, state as described in the previous section we are in a position to explain how a potential cloning paradox in de Sitter is avoided. Obviously, this will again be analogous to the AdS eternal black hole scenario described in \cite{Maldacena:2017axo}. If the exchange of information between Alice and Bob is interpreted in terms of decoding Hawking modes, which is equivalent to a gravitational shockwave procedure, it is clear that the message can never be intersected twice and a cloning paradox will be avoided.   

\begin{figure}[h]
\centering
\includegraphics[scale=1]{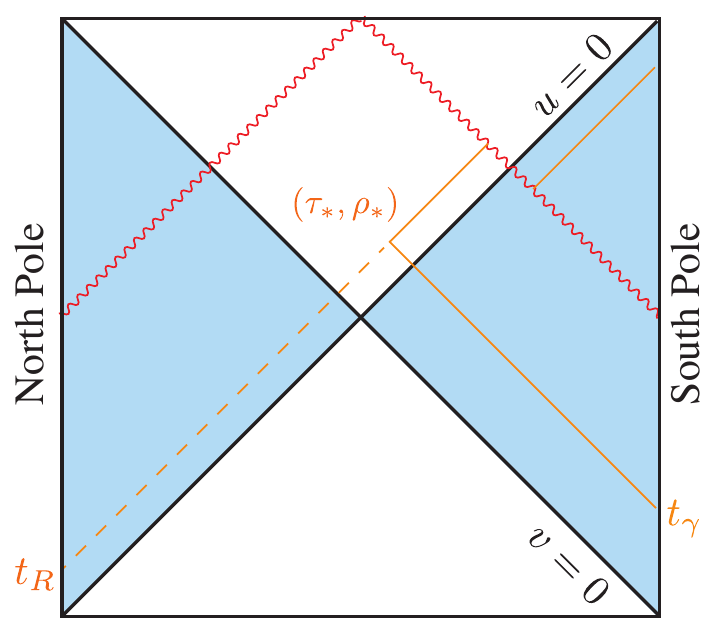}
\caption{A message leaving the South Pole is reflected and after crossing the shockwave returns to the pole.}
\label{fig:dS-reflected}
\end{figure}

To apply this mechanism restricted to just one of the poles, let us say Bob at the South pole, we will be forced to introduce the ad-hoc reflection of the actual message as soon as it has passed the cosmological horizon. Again, Bob could naively, after having collected enough Hawking modes and decoding the information in the message, decide to jump after and directly capture the message, potentially observing the same bits twice. Using the global null coordinates $u$ and $v$ we will in particular try to quantify the dependence on the reflection time. We will introduce isotropic shockwave shell perturbations at $t=0$ and a null ray message that is emitted $N_\gamma>0$ e-folds {\it before} $t=0$ from the South Pole at a fixed $v_\gamma=\ell e^{-N_\gamma}$. This null ray is reflected after having crossed the horizon to a null ray at fixed (and positive) $u_R$, where $u_R$ should be smaller than $\ell$ to allow the reflected null ray to either intersect the isotropic shockwave shell emitted from the South Pole, or a South Pole observer jumping after it, at $t=0$. Let us identify the reflected null ray as $u_R=\ell e^{-\sigma_R}$, where $\sigma_R=t_R/\ell$ should be positive in order for the reflected null ray to be potentially intercepted. To be explicit, after having introduced the null coordinates of the emitted and reflected message, we can now identify the location and time of the reflection event (in planar coordinates $\tau$ and $\rho$)
\begin{align}
e^{-\tau_*/\ell} &= \frac12e^{-\sigma_R}\left(e^{N_\gamma+\sigma_R}-1\right) ~, \\
\rho_*/\ell &= \frac12 e^{-\sigma_R}\left(e^{N_\gamma+\sigma_R}+1\right) ~.
\end{align}
The geometry of this specific one-sided protocol is sketched in Figure \ref{fig:dS-reflected}.
Note that the expression for time $\tau_*$ implies that the time difference, measured in e-folds $\Delta N_*$, between emission from the South Pole and reflection equals 
\begin{equation}
    e^{\Delta N_*}=\frac{2 e^{(\sigma_R+N_\gamma)}}{e^{(\sigma_R+N_\gamma)}-1} \, .
\end{equation}
So the typical number of e-folds after which the message is reflected is approximately $\log 2$, when $e^{(\sigma_R+N_\gamma)} \gg 1$. In that typical scenario the location of the reflection event is indeed, as expected, very close to the horizon. The only way to avoid this is to carefully fine-tune and send the message just before $t=0$ and reflect just before the maximum value of $u_R=\ell$, when both $N_\gamma$ and $\sigma_R$ are small. This just confirms the intuition that the message, for $N_\gamma \gg 1$, has to be reflected immediately after crossing the horizon.

Having defined the reflected null curve $u_R = \ell e^{-\sigma_R}$, the message will intersect with the isotropic null shell for any $\sigma_R>0$ ($\sigma_R \rightarrow \infty$ corresponding to the limit that the message is travelling along the future horizon). As we saw, the shift produced by the intersecting shell is given by $N_s \ell/(2S_{\rm dS})$. Thus, after the message hits the shockwave the null curve will be shifted to  
\begin{equation}
    {u_\gamma}^\prime =u_\gamma - \alpha = \ell\left( e^{-\sigma_R} - \frac{N_\gamma}{2 S_{\rm dS}} \right) \, ,
\end{equation}
where we assumed that the shockwave is formed by collecting Hawking quanta for $N_\gamma$ e-folds. For the message to move into the South Pole static patch region ${u_\gamma}^\prime$ should become negative. It is easy to show that this can always be realized for large enough $\sigma_R$ (close enough to the future horizon). For a given $N_\gamma$, we find the following bound
\begin{equation}
    \sigma_R > \log{\left( \frac{2 S_{\rm dS}}{N_\gamma}\right) } \, .
\end{equation}
implying that the reflection occurs at distances $\lesssim N_\gamma/M_p$ away from the South Pole horizon.

Finally, we consider the time difference between the message leaving the South Pole at $N_\gamma>0$ e-folds before $t=0$ and the time it returns to Bob at the pole. This number of e-folds is given by
\be
\Delta N_{\rm recover} =N_\gamma - \log\left(\frac{N_\gamma }{2 S_{\rm dS}} - e^{-\sigma_R}\right) ~.
\ee
For Bob to get back the message as fast as possible, he can extremize with respect to $N_\gamma$ to find
\be
\Delta N_{\rm recover} = 1 + 2 e^{-\sigma_R} S_{\rm dS} + \log(2 S_{\rm dS}) ~.
\ee
Thus, recovery happens as fast as possible when $\sigma_R \gg 1$ in which case $\Delta N_{\rm recover} \simeq \log(2 S_{\rm dS})$. As expected, the fastest possible recovery time has the same logarithmic scaling with the entropy as the scrambling time.

However, shifting the message back into the South Pole static patch is not guaranteed. For a given $N_\gamma$ the shift can be too small, or equivalently the reflection could happen too late, to map the message back into Bob's causal region. Only when introducing sufficiently energetic null shells at $t=0$, and reflection close enough to the horizon, can the information be returned back into Bob's causal region. In this particular setup, which is incomplete as a description of information recovery due to the reliance on the reflection time, this procedure removes the message from behind the horizon back into Bob's causal region. If Bob (or a surrogate) then decides to jump after the message, he will obviously not intersect the message twice. If instead Bob does not invoke the shockwave protocol, he can attempt to capture the message by chasing after it. Clearly there is never a second copy of the information, avoiding a cloning paradox. Note that the dependence on the reflection time is absent when we instead interpret the successful exchange of information between antipodal observers in terms of the decoding of Hawking radiation.

We want to stress that the derived constraints and the appearance of the scrambling time are consistent with an interpretation of the shockwave procedure in de Sitter as a decoding protocol for the cosmological Hawking radiation, but it clearly does not constitute a proof. As in the AdS eternal black hole case, it certainly seems plausible that a de Sitter bulk decoding protocol would physically move the message, due to gravitational backreaction, but strictly speaking we have only shown this for the exchange between antipodal observers. Nevertheless, we believe the above example supports the idea that a successful decoding protocol in de Sitter space physically removes the message from behind the horizon, implying the absence of a second copy, avoiding a cloning paradox.

\section{Conclusions}

In this paper we studied the use of (isotropic) shockwaves in transferring information between antipodal observers in de Sitter space and a potential interpretation of this procedure. Although our focus was on three-dimensional isotropic de Sitter shockwaves, we expect the results to generalize straightforwardly to four and higher dimensions. We first derived a general bound on the amount of information that can be transferred in terms of the number of `classical bits' required to set-up the shockwave. This immediately introduces the de Sitter entropy as providing an absolute upper bound on the amount of information that can be transferred, as expected, instead of relying on the introduction of additional species, as in earlier works.  In addition, we showed that the amount of information has to be smaller than the energy of the shockwave in units of the de Sitter temperature. This is certainly what is expected if the exchange can be understood in terms of a quantum teleportation protocol in a putative (holographically) dual description. The particular scaling with the square of $N_s$, as a consequence of the probe approximation, would be interesting to understand from a quantum information, holographic, point of view. 

Finally, we observed that the scrambling time appears naturally as the minimal amount of time one has to wait before the information between antipodal observers can be exchanged. All these properties are consistent with, and clearly suggestive of, an interpretation in terms of a (microscopic, quantum) Hayden-Preskill decoding protocol. If indeed the successful exchange of information is interpreted as an effective bulk gravitational description of the decoding of de Sitter Hawking radiation, analogous to what has been proposed for the AdS eternal black hole, a cloning paradox is obviously avoided and it clearly suggests the existence of a holographic description in terms of an entangled thermofield double construction, associated to any (antipodal) pair of free-falling observers in de Sitter space.  

We also stressed that in order to be able to interact with the cosmological horizon
an energy reservoir is needed, which can be prepared by collecting cosmological Hawking radiation, implying that the reservoir will be entangled across the two antipodal observers, preparing a symmetric, finite-energy state. In four dimensions or higher this energy reservoir can be collapsed into a black hole, producing a shared interior between the antipodal observers, in addition to the shared exterior de Sitter region. This additional Einstein-Rosen bridge can likely be manipulated as well to transfer information between the two antipodal observers. The precise role of this additional information channel in de Sitter space deserves further study, and we hope to come back to it in future work. Dispersing all of the available energy in this state through a symmetric pair of isotropic shockwave shells is then understood as a (boost symmetric) transition between the Schwarzschild-de Sitter state and the empty (Hartle-Hawking) de Sitter state, temporarily opening up a wormhole between the two antipodal static patches.  

We should stress that we have not constructed a bulk protocol involving isotropic shockwaves that can describe information recovery in a single static patch, at least not without making additional assumptions. For instance, if the antipodal region can somehow be identified as an exact copy, even after the preparation of the maximally entangled Schwarzschild-de Sitter state, then the information expelled through the cosmological horizon can possibly be returned through a traversable wormhole connecting the energy reservoirs. This would in fact be analogous to the effective interpretation of information recovery for AdS black holes, close to the thermofield double state. We hope to investigate this further, but for now we think the proposed shockwave protocol for information exchange in de Sitter space, relying on the availability of an entangled energy reservoir, provides an interesting first step towards understanding information recovery in de Sitter space and the associated nature of the de Sitter entropy. 
In that regard it would clearly be of much interest to connect our results to recent works using fine-grained entropy as a probe of quantum information. In particular, quantum extremal islands (see \cite{Almheiri:2020cfm,Marolf:2020rpm,Raju:2020smc} for reviews) have proven to be a crucial ingredient in recovering information. Finally, we should point out that the construction of de Sitter shockwaves and the possibility to use them for information exchange was also explored in the thesis work \cite{Verheijden:2018}, as well as a set of unpublished notes, of E. Verheijden. We are looking forward to follow up and report on further progress on these issues in the near future. 

\acknowledgments

We thank Erik Verlinde for interesting discussions and especially Evita Verheijden for bringing to our attention her earlier work on this subject and suggesting some additional references. This work is supported in part by the DOE under grant DE-SC0017647, the National Science Foundation under Grant No. NSF PHY-1748958, and the Kellett Award of the University of Wisconsin. We gratefully acknowledge the hospitality of the Kavli Institute for Theoretical Physics during the workshop ``The String Swampland and Quantum Gravity Constraints on Effective Theories'' where part of this work was completed. AC thanks the Kavli Institute for Theoretical Physics for a Graduate Fellowship during this period. This work is part of the Delta ITP consortium, a program of the Netherlands Organisation for Scientific Research (NWO) that is funded by the Dutch Ministry of Education, Culture and Science (OCW). EM is supported by Ama Mundu Technologies (Grant Adoro te Devote).

\vspace{1cm}

\appendix

\section{De Sitter invariance of stress tensor in Bunch-Davies state} \label{app:reduceddensitymatrix}
In this appendix, we confirm the expectation that the vacuum expectation value of the stress tensor in the (pure) Bunch-Davies state computed using the (mixed) reduced density matrix respects de Sitter invariance. For simplicity we work in two dimensions, but our results can be easily generalized to higher dimensions. First, we define a set of null coordinates $(U,V)$ that cover the static patch surrounding the South Pole. With respect to the global $(u,v)$ coordinates they are defined as
\be
u = -\ell e^{-U/\ell} ~, \quad v = \ell e^{V/\ell} ~.
\ee
Considering a massless scalar field $\hat\varphi$, we can expand in modes that are positive frequency with respect to these coordinates.
\be \label{eq:staticmodeexpansion}
\hat\varphi = \int_0^\infty \frac{dk}{\sqrt{4\pi k}}\left[e^{-ik U}\hat b_k + e^{+ik U}\hat b^\dagger_k \right] + (U \leftrightarrow V) ~.
\ee
The creation/annihilation operators obey
\be
[\hat b_k,\hat b_{k'}^\dagger] = \delta(k-k') ~.
\ee
The vacuum state that is natural for a static observer at the South Pole is then defined as
\be
\hat b_k \ket{0}_S= 0 ~.
\ee
The Bunch-Davies vacuum is defined by expanding with respect to a different set of mode functions as
\be
\hat\varphi = \int_0^\infty \frac{dk}{\sqrt{4\pi k}}\left[e^{-ik u}\hat a_k + e^{+ik u}\hat a^\dagger_k \right] + (u \leftrightarrow v) ~,
\ee
and imposing
\be
\hat a_k\ket{0}_{\rm BD} = 0 ~.
\ee
Using the Bogoliubov transformation between both sets of creation/annihilation operators
\be
\hat b_k = \int_0^\infty dk'\left(\alpha_{kk'}\hat a_{k'}+\beta_{kk'}\hat a^\dagger_{k'}\right) ~,
\ee
with 
\be
|\beta_k|^2 = \frac1{e^{2\pi k\ell}-1} ~,
\ee
and $|\alpha_k|^2-|\beta_k|^2=1$ we can write the Bunch-Davies state as
\be
\ket{0}_{\rm BD} = \prod_k \frac1{|\alpha_k|}\sum_n \left(\frac{\beta_k}{\alpha_k}\right)^n \ket{n_k}_S ~.
\ee
Then, the reduced density matrix at the South Pole is obtained by tracing over the North Pole.
\be
\rho_S \equiv \text{tr}_N\left(\ket{0}_{\rm BD}\bra{0}_{\rm BD}\right) = \prod_k(1- e^{-2\pi k\ell}) \sum_n e^{-2\pi k n\ell}\ket{n_k}_S\bra{n_k}_S ~.
\ee
The vacuum expectation value of the stress tensor is now given by
\be
\braket{T_{ab}}_{\rm BD} = \text{tr}_S(T_{ab}\rho_S) ~.
\ee
Using the expansion in modes \eqref{eq:staticmodeexpansion} that are only defined in a single static patch, we find that the diagonal components of the stress tensor are given by
\begin{align} \label{eq:diagonalstress}
\braket{T_{UU}}_{\rm BD} = \braket{T_{VV}}_{\rm BD} = \int_0^\infty \frac{dk\, k}{4\pi}\left(1+\frac{2}{e^{2\pi k\ell}-1}\right) ~. 
\end{align}
The off-diagonal component is fixed by the (state-independent) conformal anomaly
\be
\braket{T^a_{\,\,\,a}} = \frac{1}{12\pi\ell^2} ~.
\ee
We recognize the first term in \eqref{eq:diagonalstress} as a UV-divergence that coincides with the standard divergence in flat space. Subtracting it (which is equivalent to normal ordering with respect to $\hat a_k$) and evaluating the remaining integral we find
\be
\braket{T_{UU}}_{\rm BD} = \braket{T_{VV}}_{\rm BD} = \frac1{48\pi\ell^2} ~,
\ee
which is the stress tensor of a two-dimensional boson gas at inverse temperature $\beta = 2\pi\ell$ \cite{Aalsma:2021bit}. Note that although the stress tensor expressed in this coordinate system is not proportional to the metric, this does not imply backreaction as normal ordering breaks general covariance \cite{Fabbri:2005mw}.

We can now transform this result to the globally defined lightcone coordinates using the transformation law
\be
\braket{T_{UU}} = (u')^2\braket{T_{uu}} -\frac{1}{24\pi}\{u,U\} ~,
\ee
where the second term is the Schwarzian derivative
\be \label{eq:trafolaw}
\{u,U\} = \frac{{u}'''}{{u}'} - \frac32\left(\frac{{u}''}{{u}'}\right)^2 ~,
\ee
and the prime denotes a derivative with respect to $U$. The expression for the $VV$ component of the stress tensor is simply obtained by exchanging $U\to V$. Using \eqref{eq:trafolaw} we can express the stress tensor in $(u,v)$ coordinates.
\be \label{eq:stressfinal}
\braket{T_{ab}}_{\rm BD} = \frac{1}{24\pi\ell^2}g_{ab} ~.
\ee
Clearly, this result is manifestly de Sitter symmetric. Finally, we could transform this to planar coordinates, as was done in \cite{Markkanen:2017abw}, which results in the same expression \eqref{eq:stressfinal} because the Schwarzian term vanishes for this transformation. Our de Sitter invariant result for the stress tensor disagrees with Eq. (76) and (77) of \cite{Markkanen:2017abw}, which was obtained by (incorrectly) omitting the Schwarzian when transforming between coordinates defined in the static patch and planar coordinates.

Although we only considered the two-dimensional case in detail here the same logic applies to the four-dimensional case. Tracing over an unobservable region cannot break the de Sitter symmetries and the (renormalized) stress tensor computed using the reduced density matrix will be proportional to the metric. In higher dimensions, additional care must be taken in the renormalization procedure as it is possible to draw the (incorrect) conclusion from the unrenormalized stress tensor that the vacuum expectation value will not respect the symmetries of the problem, as was explained elegantly in a recent work \cite{Kourkoulou:2021ksw}.

\bibliographystyle{JHEP}
\bibliography{refs}

\end{document}